\documentclass[reprint,amsmath,amssymb,aps,superscriptaddress]{revtex4-2}

\usepackage{graphicx}
\usepackage{dcolumn}
\usepackage{bm}
\usepackage{verbatim}
\usepackage{xcolor}
\usepackage{subfig}
\usepackage{lipsum}

\begin{document}

\title{Deep Learning for Optical Tweezers}

\author{Antonio Ciarlo}
\email{antonio.ciarlo@physics.gu.se}
\affiliation{University of Gothenburg, 
Department of Physics, Gothenburg, Sweden}

\author{David Bronte Ciriza}
\affiliation{CNR-IPCF, Istituto per i Processi Chimico-Fisici, Messina, Italy}

\author{Martin Selin}
\affiliation{University of Gothenburg, 
Department of Physics, Gothenburg, Sweden}

\author{Onofrio M. Maragò}
\affiliation{CNR-IPCF, Istituto per i Processi Chimico-Fisici, Messina, Italy}

\author{Antonio Sasso}
\affiliation{Università degli studi di Napoli Federico II, Dipartimento di Fisica ``Ettore Pancini'', Naples, Italy}

\author{Giuseppe Pesce}
\affiliation{Università degli studi di Napoli Federico II, Dipartimento di Fisica ``Ettore Pancini'', Naples, Italy}

\author{Giovanni Volpe}
\email{giovanni.volpe@physics.gu.se}
\affiliation{University of Gothenburg, 
Department of Physics, Gothenburg, Sweden}

\author{Mattias Goksör}
\affiliation{University of Gothenburg, 
Department of Physics, Gothenburg, Sweden}
             
\begin{abstract}
Optical tweezers exploit light--matter interactions to trap particles ranging from single atoms to micrometer-sized eukaryotic cells. For this reason, optical tweezers are a ubiquitous tool in physics, biology, and nanotechnology.
Recently, the use of deep learning has started to enhance optical tweezers by improving their design, calibration, and real-time control as well as  the tracking and analysis of the trapped objects, often outperforming classical methods thanks to the higher computational speed and versatility of deep learning.
Here, we review how deep learning has already remarkably improved optical tweezers, while exploring the exciting, new future possibilities enabled by this dynamic synergy. 
Furthermore, we offer guidelines on integrating deep learning with optical trapping and optical manipulation in a reliable and trustworthy way.
\end{abstract}
\maketitle

\section{Introduction}

Optical trapping and optical manipulation exploit light--matter interactions to trap and manipulate various types of micro- and nanoparticles.
These techniques date back to Arthur Ashkin, who demonstrated in the 1970s that it is possible to levitate microparticles in a fluid using a focused laser beam~\cite{ashkin1970acceleration, ashkin1970atomic, ashkin1977feedback, ashkin1986observation}. Later, A. Ashkin and coworkers  demonstrated that it is also possible to trap particles in 3D using a strongly focused laser beam~\cite{ashkin1986observation}  --- a technique now known as optical tweezers~\cite{jones2015optical,volpe2022roadmap}.
 
Optical tweezers are now an ubiquitous tool in science, allowing for flexible, non-invasive manipulation of nano- and micro-particles as well as for the measurement of forces acting on them.  
Both trapping and force measurement using optical tweezers have proved fundamental in fields ranging from statistical mechanics~\cite{mccann1999thermally,bechinger2001phase,ciarlo2023fickian,pastore2022model,pastore2021rapid}, nanothermodynamics~\cite{gieseler2018levitated}, soft matter~\cite{lowen2001colloidal,petrov2007raman} and biology~\cite{bustamante2021optical,block1990bead,finer1994single} to microfabrication~\cite{holmlin2000light,agarwal2005manipulation} and atomic physics~\cite{grimm2000optical,gustavson2001transport,meschede2006manipulating}. Different kinds of optical tweezers have been developed to tackle the specific challenges of each application, such as trapping of nanoparticles using plasmons~\cite{volpe2006surface, juan2011plasmon} and Raman tweezers~\cite{petrov2007raman}. 

Deep learning is a collection of computer algorithms that can improve and adapt their solutions by learning the rules connecting input and output directly from data~\cite{lecun2015deep}, solving problems ranging from particle tracking and characterization~\cite{midtvedt2021quantitative} to protein folding~\cite{alphaFold2021highly} and face recognition~\cite{balaban2015deep}. The first steps towards the deep learning revolution were taken in the 1940s with the mathematical modeling of biological neurons by neuroscientist Warren McCulloch and logician Walter Pitts~\cite{mcculloch1943logical}. The recent growth of deep learning has been driven largely by the recent increase in the computational power of processors and the size of datasets, but also by the spread of user-friendly all-purposes deep learning frameworks, such as PyTorch~\cite{paszke2017automatic,NEURIPS2019_9015} and Keras/Tensorflow~\cite{tensorflow2015-whitepaper,chollet2015keras}, which enable quick and easy deployment of deep learning solutions for a wide range of tasks. 

Several aspects of optical tweezers that are difficult to study theoretically, either due to the computational cost or because of the high modelling complexity, can now be addressed using deep learning. 
Deep learning can improve the calculation of optical forces by increasing its speed~\cite{lenton2020machine} and even accuracy~\cite{bronte2022faster}, helping to realistically simulate more complex systems. From an experimental standpoint, deep learning can enhance the calibration of optical tweezers~\cite{argun2020enhanced} and improve the tracking of trapped particles~\cite{helgadottir2019digital}. Furthermore, recent progress in deep learning is also benefiting the real-time control of optical tweezers~\cite{aggarwal2010real} and the design optimization~\cite{li2019algorithmic}. 

This review presents an overview of optical tweezers and deep learning, highlighting their recent collaborative developments. We speculate on possible future innovations resulting from this synergy. To conclude, we suggest strategies for those aiming to utilize deep learning in combination with optical trapping and optical manipulation in a reliable and safe way.

\section{Optical tweezers}

\begin{figure*}
    \centering\includegraphics[width=\textwidth]{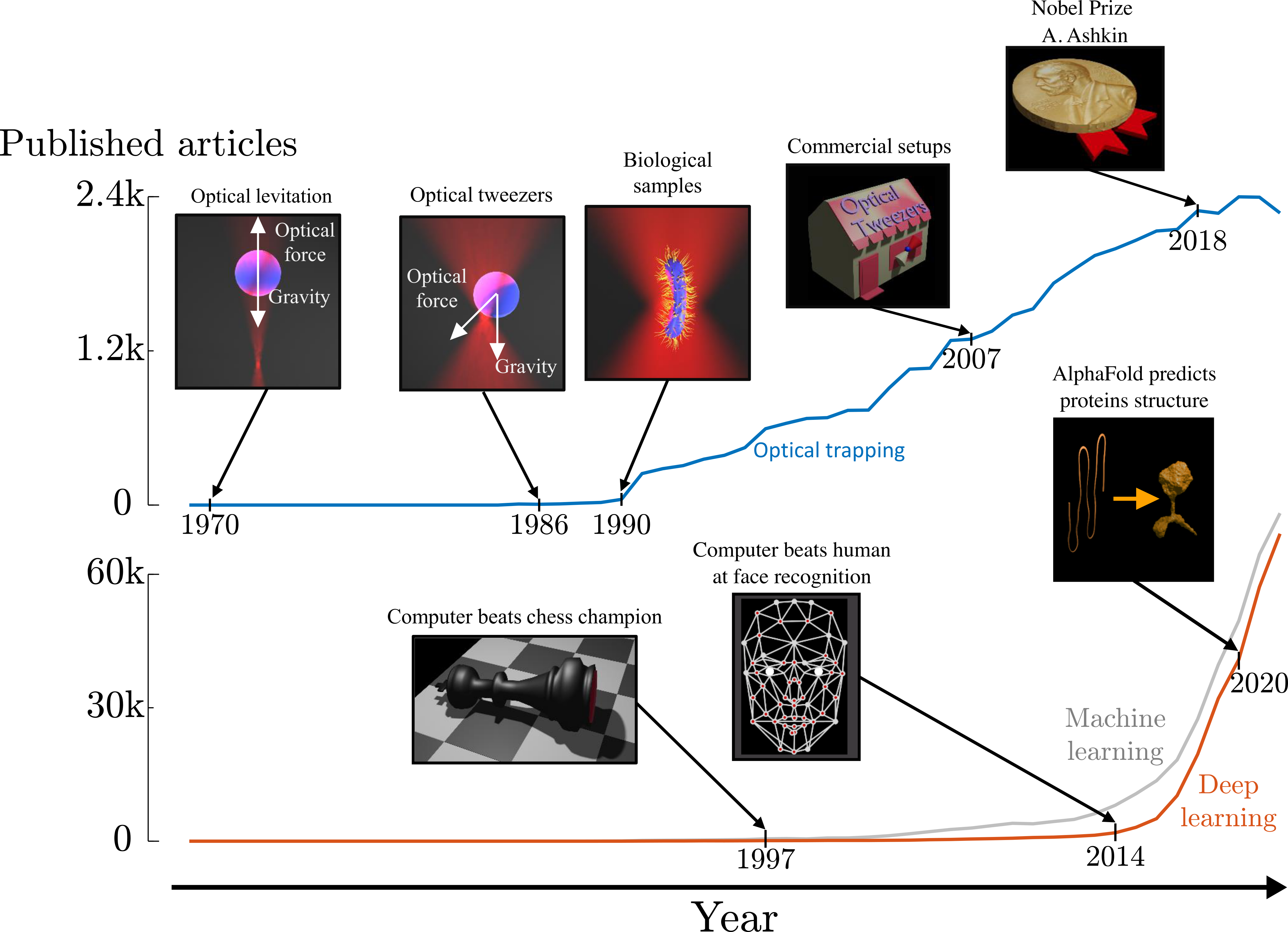}
    \caption{
    {\bf The rise of optical trapping and deep learning in scientific publications.}
    Number of articles published per year that use ``Optical trapping'' (blue line),``Machine learning'' (gray line), or ``Deep learning'' (orange line) in their title, abstract, or keywords. Milestones in the development of these fields are highlighted with illustrations. Data obtained from Web of Science\texttrademark on November 2023.}
    \label{fig:papers}
\end{figure*}

Optical tweezers are an ubiquitous tool in science and they are contributing to the progress of fields like biology, physics, and nanotechnology~\cite{jones2015optical,volpe2022roadmap}. As can be seen in Fig.~\ref{fig:papers}, the field of optical trapping and optical manipulation is rapidly expanding.
Based on light--matter interactions, optical forces can trap particles in the proximity of a focused laser beam. Furthermore, the trapping forces are typically so small that, by employing a trapped particle as a probe, it is possible to measure forces well below those reachable with an atomic force microscope (AFM) and micro-fabricated cantilevers~\cite{binning1986AFM}.
Despite of the recent progress in optical trapping, there are still many open challenges~\cite{volpe2022roadmap}, including the calculation of optical forces, the efficient calibration of an optical trap, the position detection of a trapped particle, and the development of new optical trapping systems.

The calculation of optical forces has typically relied on approximations that depend on the trapping regime defined by size of the particle~\cite{jones2015optical, pesce2020optical}. The trapping regimes are the geometrical-optics regime, the Rayleigh regime, and the intermediate regime.
The \emph{geometrical-optics regime} is valid when the size of the particle is much larger than the wavelength $\lambda_0$ of the trapping light. In this case, the wave nature of the light can be neglected and optical forces can be calculated using ray optics~\cite{ashkin1992forces, callegari2015computational}.
Instead, the \emph{Rayleigh regime} occurs when the linear dimensions of the trapped object are much smaller than $\lambda_0$. Thus, the trapped object behaves like a dipole and the optical forces are mostly proportional to the gradient of the light intensity~\cite{chaumet2000time}.
Finally, the \emph{intermediate regime} lays in between, where the linear dimensions of the trapped object are comparable with $\lambda_0$. In this case, the optical forces need to be calculated from the electromagnetic fields obtained as an exact solution of the scattering problem, which can be a very complex and computationally intensive process~\cite{borghese2007scattering,mishchenko2006multiple,nieminen2007optical}.
Common to all the regimes is that the trapping forces for small displacements from the trapping position can be approximated as an harmonic force
\begin{equation}
    F(r)=-k\cdot r,
\end{equation}
where $k$ is the stiffness of the trap, $r$ is the displacement from the equilibrium position, and $F(r)$ is the optical force.

Calibrating an optical tweezers consists of determining the relation between the position of a particle and the force it experiences.
For small displacements from the equilibrium position, it is sufficient to determine the trap stiffness. The traditional approaches to calibration rely on explicit mathematical recipes such as the potential method~\cite{florin1998photonic}, the autocorrelation method~\cite{viana2002dynamic}, the power spectrum analysis~\cite{berg2004power}, the mean square displacement method, the equipartiton method, or the maximum-likelihood-estimator analysis (FORMA)~\cite{perez2018high}. While these approaches perform well when the field is static, conservative, and a high amount of data are available, they present some limitations when the force field does not satisfy these assumptions. 

In optical trapping experiments, the location of the particle is often the most critical parameter. Even though the previously mentioned calibration techniques differ in their approaches, they all rely on this knowledge.
There are two main possibilities for tracking the position of the particle. For a single particle in an optical trap, one can use the trapping laser as a probe to determine its position, for instance using a quadrant photodiode (QPD) or a position sensitive detector (PSD).
However, when there are multiple particles or multiple traps, interpreting the QPD signal becomes more complex and cameras are typically necessary. These cameras provide a larger view of the experimental system under investigation, containing much more information than the QPD/PSD signals but with the drawback of a lower acquisition rate.

Nowadays, in order to expand the applicability of optical trapping, new techniques to control optical tweezers are being developed. External real-time feedback allows to correct the trapping force by adjusting either the intensity of the light or the position of the trap~\cite{simmons1996quantitative, wallin2008stiffer}. Introducing external feedback increases the effective trap stiffness but comes with the drawbacks of a limited bandwidth and of a higher sensitivity to errors in the detection of the position of the particle. To overcome these problems, automatic feedback control mechanisms have been postulated for plasmonic tweezers~\cite{juan2011plasmon} and realized for intracavity optical trapping~\cite{kalantarifard2019intracavity}.

\section{Deep learning}

\begin{figure*}
    \centering
    \includegraphics[width=\textwidth]{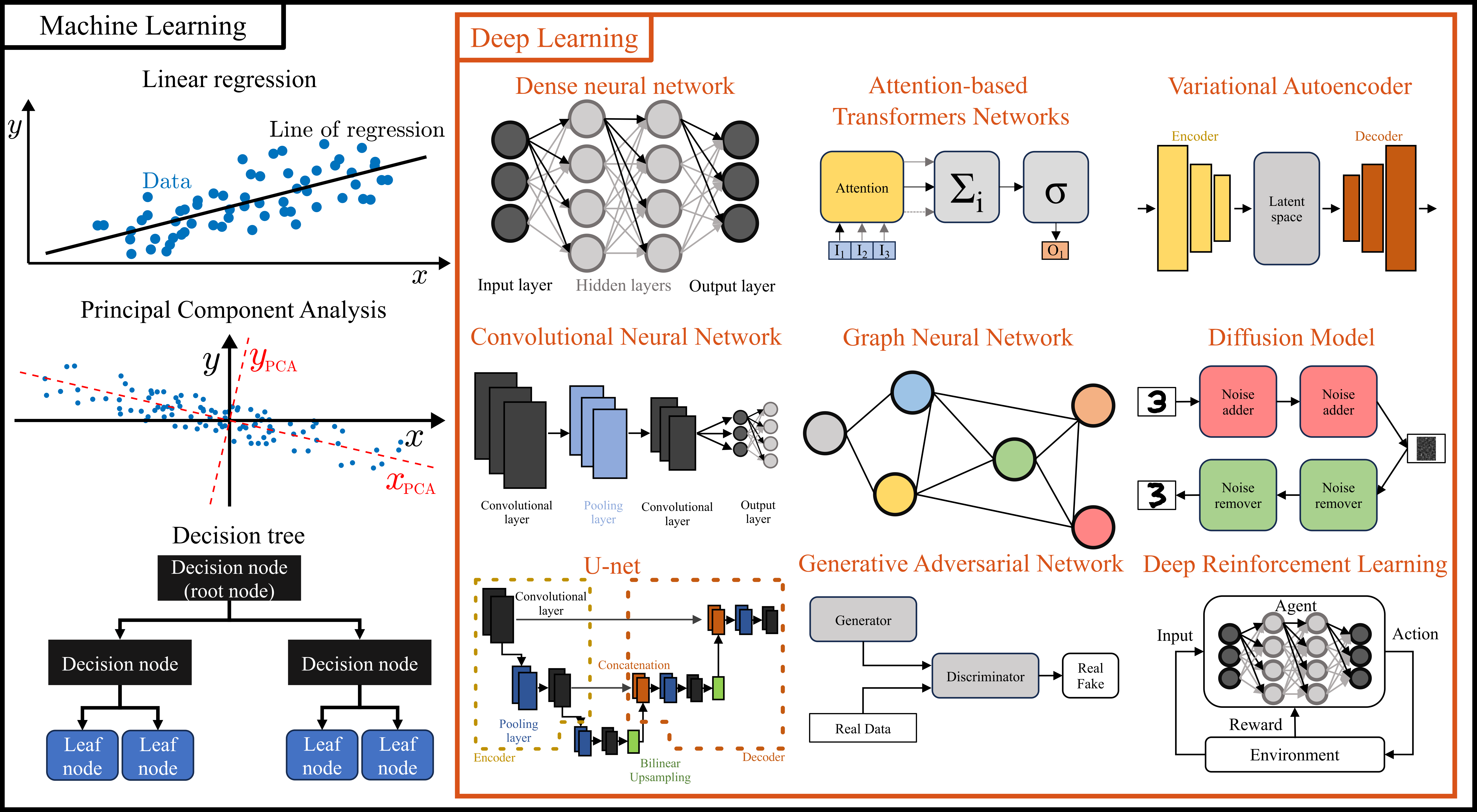}
    \caption{{\bf Machine learning and deep learning.} 
    Deep learning (orange rectangle) is a subset of machine learning (black rectangle). 
    Machine learning approaches include linear regression, principal component analysis, and decision trees. 
    Deep learning approaches include dense neural networks, convolutional neural networks, U-nets, attention-based transformer networks, graph neural networks, generative adversarial networks, variational autoencoders, diffusion model, and deep reinforcement learning.}
    \label{fig:MachineLearningAlgorithms}
\end{figure*}

Deep learning is a branch of computer science that, by using artificial neural networks, allows computers to learn from data and improve their performance without explicit programming.
It is a subset of machine learning, as shown Fig.~\ref{fig:MachineLearningAlgorithms}. Typically, deep learning approaches extract hierarchical features from data to realize complex tasks such as image recognition, natural language processing, and speech synthesis with remarkable accuracy and efficiency. They achieve this by automatically learning hierarchical features from raw data, reducing the need for manual feature engineering, whereas traditional machine learning models, like linear regression, principal component analysis, or decision trees, often require explicit feature extraction. This has led to a near-exponential growth in the use of machine learning and, in particular, deep learning~\cite{lecun2015deep}, as shown in Fig.~\ref{fig:papers}.

Deep Learning is typically based on deep (i.e., multi-layer) artificial neural networks with many trainable parameters that transform input data into output data~\cite{lecun2015deep}. 
These parameters are automatically adjusted during the \emph{training process}, in which the system learns the rules that connect the input data to the desired outputs by operating on known input/output pairs, called training data, using algorithms such as stochastic steepest descent and error backpropagation~\cite{rumelhart1986learning}.
Thus, specific problems can be addressed reliably without explicitly knowing the rules connecting input and output, especially when the data to be analyzed closely resemble the training data.

The fundamental building block of neural networks is the artificial neuron~\cite{mcculloch1943logical}. 
The artificial neuron processes its inputs by  performing a weighted sum and returning a transformation (typically a nonlinear activation function) of the resulting sum. During the training process, the trainable parameters, often referred to as weights, are tuned to optimize the output of the neuron. Artificial neurons can be connected in layers, with each neuron receiving input from neurons of the previous layer and passing its output to the next layer, forming the most standard artificial neural network.

Deep learning can be implemented through different network structures, i.e., different architectures, and choosing the right one depends on the task at hand. 
The efficiency and effectiveness of the solution are strongly influenced by the architecture because different problems have distinct data characteristics and complexities. Generally speaking, more complex data require more complex models in terms of the number of parameters needed for fitting and analysis. Moreover, deep learning can be used to generate synthetic data of high quality. 
For our purposes, it is convenient to group the different architectures into three groups based on their purposes: data analysis (Dense Neural Networks, Convolutional Neural Networks, U-nets, Recurrent Neural Networks, Transformers Networks, Graph Neural Networks), data generation (Generative Adversarial Networks, Variational Autoencoders, Diffusion Models), and decision making (Deep Reinforcement Learning).

\subsection{Data analysis} \label{sec:data_analysis}

Dense neural networks (DNNs) are artificial networks in which all the nodes in each layer are connected to all the nodes in the adjacent layers. 
They have a structure characterized by a first layer referred to as the input layer and a last layer referred to as the output layer (dark circles in Fig.~\ref{fig:MachineLearningAlgorithms}), and one or more layers in between referred to as hidden layer (gray circles in Fig.~\ref{fig:MachineLearningAlgorithms}). They are used to deal with tabular data, sequential data, and data with small dimensions. When dealing with high-dimensional data, such as images, the number of connections between the layers increases drastically leading to problems such as overfitting, meaning that the neural network performs exceptionally well on the training data but fails to generalize to new, unseen data.

To deal with high-dimensional data, convolutional neural networks (CNNs) employ 2D layers of neurons partially connected one to the other~\cite{fukushima1980neocognitron,lecun1998gradient,krizhevsky2012imagenet}.
The key layers are the convolutional layers, which use filters to scan the input and perform convolutional operations, as shown in Fig.~\ref{fig:MachineLearningAlgorithms}. A filter uses the same weights for different subsets of the input image, thus reducing the number of required trainable parameters and the risk of overfitting. More importantly, each filter corresponds to a feature map that detects a feature in the input data. In this way, the convolutional layer can detect different features of the input for each of its filters. Typically, the image size decreases as it passes through the layers, reducing the computational load and providing access to the information present at different length scales. Often, a dense neural network is added to the final layer of the convolutional neural network to generate an output representing comprehensive information associated with the input, for example, the coordinates of the position of a particle~\cite{helgadottir2019digitalDeeptrack}. 
By reducing the dimensionality of the input, CNNs identify more abstract and high-level features from the data, such as the general shape of a particle or cell, at the expense of low-level features.
Therefore, CNNs excel in image detection, recognition, and segmentation~\cite{chauhan2018convolutional, bullock2019xnet}.

U-nets~\cite{ronneberger2015u} are characterized by their ``U-shaped'' design consisting of a contracting path (encoder) connected to an expanding path (decoder) connected also by skip connections, as shown in Fig.~\ref{fig:MachineLearningAlgorithms}. 
These skip connections bridge earlier and later layers in the network, ensuring that both low-level and high-level features are effectively combined by enabling the direct transfer of feature maps. 
The contracting path reduces the dimension of the input thanks to several convolutional layers, capturing and summarizing local information to learn high-level features. Instead, the expanding path consists of transposed convolutions (or deconvolutions) to up-sample the feature map restoring the dimension of the input. Through the skip connections, the expanding path receives high-resolution feature maps preserving the low-level features in the final output. Between the contracting and expanding path, i.e., at the bottom of the U shape, there is a bottleneck layer having the most abstract and high-level representation of the input data. Even if U-nets solve the loss of low-level features, they still need, like any CNN, a large number of diverse training data to reach good performances and acceptable reliability. For example, U-Nets have achieved significant success in the analysis of brain tumors images from MRI scans~\cite{shokiche2016high}, denoising astronomical images~\cite{vojtekova2021learning}, and characterizing the microstructure of samples imaged with scanning electron microscopy~\cite{bangaru2022scanning}.

Unlike the previous architectures, recurrent neural networks (RNNs) retain and utilize information from previous time steps~\cite{rumelhart1985learning}. For this reason, RNNs incorporate memory gates that adjust their internal state based on prior data [55]. A fundamental characteristic of RNNs is their capability to establish recurrent connections, generating a feedback loop within the network. This enables the information to circulate within the network, making it responsive to the order and timing of input data. 
However, conventional recurrent neural networks encounter constraints resulting in difficulties in capturing prolonged dependencies effectively, including the vanishing gradient problem~\cite{mehlig2021machine}.
To address this issue, advanced models such as long short-term memory (LSTM)~\cite{hochreiter1997long} and gated recurrent unit (GRU)~\cite{cho2014learning} networks have been developed. These structures contain more advanced memory gates that can select and retain information over extended sequences, making them especially effective in tasks such as speech recognition, where long-term contextual information is crucial.
Overall, RNNs excel in applications where the sequence of data elements is important, such as natural language processing~\cite{sutskever2014sequence}, protein analysis~\cite{thireou2007bidirectional,hochreiter2007fast}, optical coherence tomography data segmentation~\cite{kugelman2018automatic}, and adaptive optics control~\cite{landman2020self}.

Attention-based transformers networks (ATNs) employ self-attention mechanisms to analyze sequential data, enabling them to identify how even distant elements in the sequence interact and influence each other~\cite{vaswani2017attention}, as shown in Fig.~\ref{fig:MachineLearningAlgorithms}.
The first step is to add some position information to the sequential input data through positional encoding (typically creating a vector applying the cosine function for every odd index of the input data and a vector applying the sine function for every even index). Then, an encoder layer maps all the input sequences into a continuous representation. It is composed of 2 sub-modules: the multi-headed attention and the dense neural network. The multi-headed attention layer allows the model to focus on specific elements of the input data, assigning them different levels of importance during the learning process thanks to a scoring matrix (determining the amount of attention one element of the input should have on the others). The word ``multi-headed'' refers to the fact that this layer analyzes simultaneously the input with a different attention sub-modules called ``heads''. The dense neural network, which follows multi-headed attention, enhances the representations of the input elements to learn higher-level information.
After the encoder, its output is sent to a decoder that has two multi-headed attention layers followed by a dense neural network. The first multi-headed attention layer receives the output of the encoder after positional encoding and sends its output to the second multi-headed layer that combines it directly with the output of the encoder (without positional encoding) allowing the decoder to understand which encoder input is relevant to put a focus on. In the end, the dense neural network classifies the input and chooses the highest probability prediction for the output. Transformers have proved themselves very useful in language modeling~\cite{devlin2018bert}, text generation~\cite{brown2020language}, and image captioning~\cite{parmar2018image}.

Graph neural networks (GNNs) are designed to analyze data organized as graphs, capturing complicated relationships within them~\cite{gori2005new, scarselli2008graph, gallicchio2010graph}, as shown in Fig.~\ref{fig:MachineLearningAlgorithms}.
A graph comprises a set of nodes (or vertices) linked by edges (or links). The nodes, in which information is stored within a vector known as a feature vector, correspond to the input data, while the edges represent the corresponding dependencies.
The process begins by taking the input graph and passing it through a sequence of neural networks. This transformation transforms the structure of the input graph into a graph embedding (i.e., into vectors), preserving essential details about nodes, edges, and overall context. Next, the feature vectors associated with the nodes are passed to a neural network layer. These features are combined and aggregated within this layer, and the resulting information is then passed on to the next layer in the network. In this way, the GNN updates node representations iteratively to capture information from neighboring nodes, often by following a series of message-passing steps. During these steps, each node aggregates information from its neighbors, applies a learnable function, and updates its representation accordingly.
The first obvious application of GNNs is the classification of nodes and the completion of graphs with missing links. More interesting applications in which GNNs excel are, for example, web recommendation systems~\cite{kipf2016semi}, traffic prediction~\cite{li2017diffusion}, protein--protein interactions~\cite{fout2017protein}.

\subsection{Data generation}

Generative adversarial networks (GANs) create high-quality synthetic data by using a specific method called adversarial training~\cite{creswell2018generative}. 
This method uses two neural networks: the generator, which produces the synthetic data, and the discriminator, which verifies whether the data are real or fake, as shown in Fig.~\ref{fig:MachineLearningAlgorithms}. The adversarial training improves the synthetic data generation by training the generator and discriminator in alternating steps. First, the generator produces synthetic data from the input data and the discriminator tries to classify them. Following this, by using both real and synthetic data, the discriminator is trained to better classify data. Finally, the generator is updated to produce more realistic data by using the results of the training of the discriminator. This adversarial process continues iteratively until the generator produces synthetic data able to deceive the discriminator. Step-by-step, the generator can produce samples that are almost indistinguishable from real data, making GANs a powerful tool in data augmentation and data synthesis applications. 
A recent evolution of GANs, called Time-series GANs (TGANs), allows the generation of time-series data by taking into account the temporal correlations of the time-series data~\cite{yoon2019time}. 
However, training GANs can be challenging because they might suffer from mode collapse (producing limited diversity in generated samples). GANs are used not only for data generation, but also for image-to-image translation~\cite{isola2017image}, for enhancing the resolution of images~\cite{ledig2017photo}, and for anomaly detection~\cite{schlegl2017unsupervised}.

Variational autoencoders (VAEs) are generative models that combine deep neural networks with probabilistic modeling to learn representations of data and generate new samples by mapping input data into a continuous latent space~\cite{kingma2013auto}. 
VAEs use deep neural networks to produce a meaningful latent space representation of the input data, where a latent space is a lower-dimensional space in which the input data are mapped into a distribution (typically, a multivariate Gaussian). To do this, VAEs use an encoder and a decoder, as shown in Fig.~\ref{fig:MachineLearningAlgorithms}. The encoder is a neural network (typically, a dense or convolutional neural network) that extrapolates from the input data the mean ($\mu$) and the variance ($\sigma$) of the distribution in the latent space. Once these two parameters are known, the encoder uses them to sample a point ($z$) from the latent space by using the reparameterization trick following a standard Gaussian distribution, i.e., $z=\mu+\epsilon \cdot\sigma$ with $\sigma$ Gaussian random noise term.
Then, the decoder uses a neural network to reconstruct the original input data from the latent space representation obtained with the encoder. In this way, it takes points from the latent space and generates a new data sample that is similar to the input data one. VAEs have proved useful to reconstruct complex many-body physics~\cite{luchnikov2019variational}, for regressions~\cite{zhao2019variational}, and for music generation~\cite{hennig2017classifying}.

Diffusion models (DMs) are a deep learning architecture created to simulate the evolving changes in data over time or space, emulating the fundamental principles of diffusion processes and allowing a heterogeneous data production~\cite{sohl2015deep}. 
These models add noise or perturbations to the input data during different steps, converting them into an uncertain state, as shown in Fig.~\ref{fig:MachineLearningAlgorithms}. Subsequently, the model is trained to reverse this process using a neural network to predict and control the noise reduction, gradually restoring the data point to its original or desired state. This approach to noise reduction produces data samples that reflect the underlying trends and variability of the data distribution while ensuring coherence, realism, and high heterogeneity thanks to the randomness of the process. This means that DMs have an exclusive ability to capture patterns and variations inherent in data distribution.
The adaptability of diffusion models cover a wide range of applications, including image generation~\cite{saharia2022photorealistic,ho2022cascaded,pinaya2022brain,kawar2023imagic} and natural language processing~\cite{liu2021deep}.

\subsection{Decision making}

Deep reinforcement learning (DRL) is a deep learning approach that combines deep neural networks with reinforcement learning techniques to learn sequential decision-making in complex environments through trial and error~\cite{mnih2013playing,mnih2015human}. 
It is based on reinforcement learning, in which an agent learns to make sequential decisions in an environment to maximize a cumulative reward signal. In DRL, the agent employs a neural network that is trained using feedback from the environment, as shown in Fig.~\ref{fig:MachineLearningAlgorithms}. This feedback consists of rewards or penalties for the agent based on its actions. Through iterative interactions with the environment, collecting experiences, and updating its neural network, the DRL agent gradually learns an optimal policy or value function, enabling it to make effective decisions in complex and high-dimensional environments. In this way, DRL can do very complex tasks like playing Go~\cite{silver2016mastering}, driving autonomous vehicles~\cite{sallab2016end}, and designing optical multi-layer thin films~\cite{wang2021automated}. 

\section{Deep learning for optical tweezers}

The advantages of machine learning, such as simplicity, versatility and speed, enhance optical tweezers by improving particle detection and tracking, trajectory analysis and calibration, optical force calculation, and by enabling tasks such as real-time control of optical traps and new designs. When automated without deep learning, these tasks typically require manual tuning of parameters, low noise measurements, or extremely long calculations. This is undesirable because it is time consuming for the researchers and also risks introducing human biases. In the following subsections, we discuss different cases where deep learning has already been successfully combined with optical trapping and optical manipulation, and we propose new possible applications.

\subsection{Particle tracking}

\begin{figure*}[t]
	\centering
	\includegraphics[width=\textwidth]{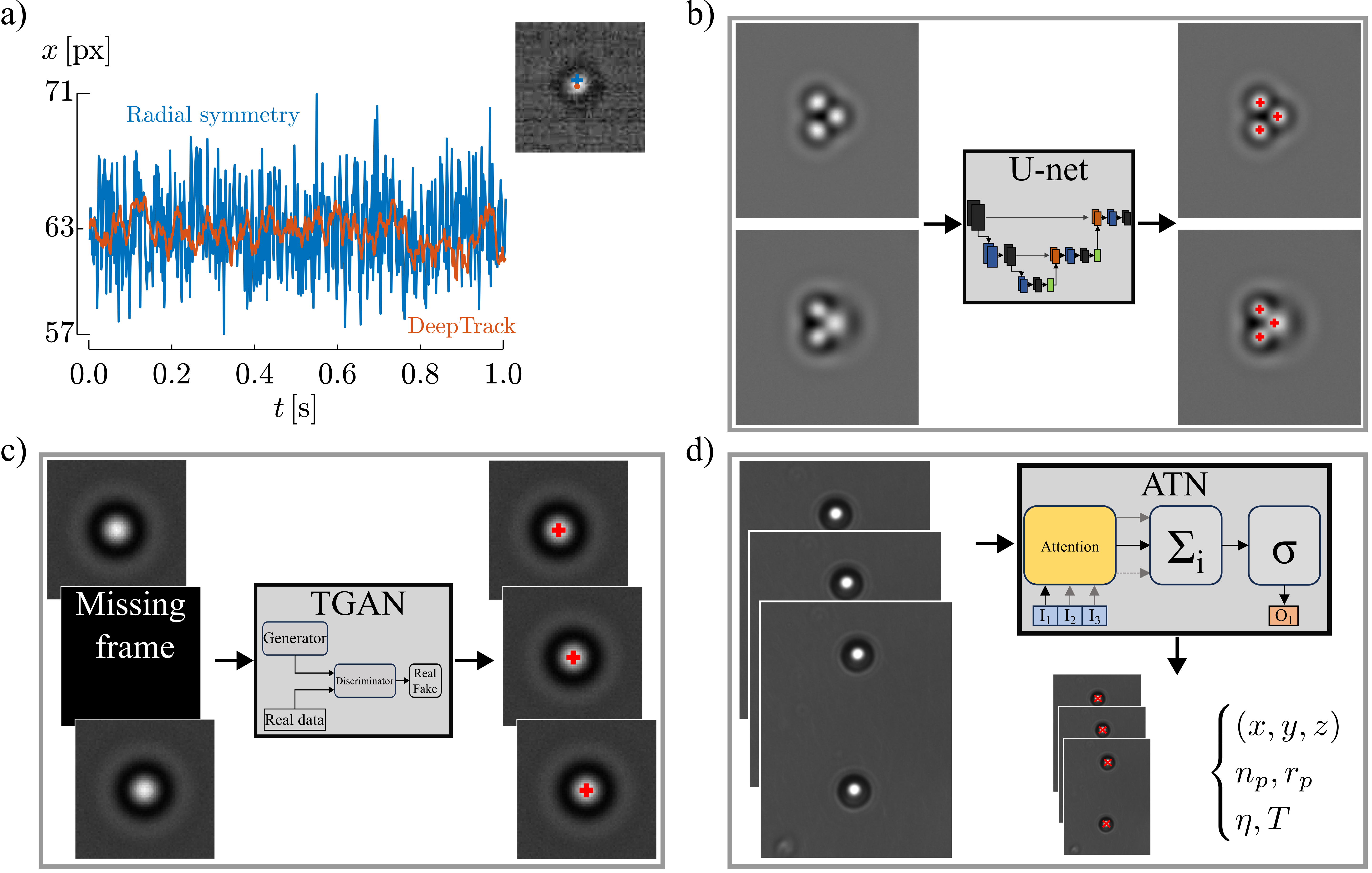}
	\caption{\textbf{Deep learning for particle tracking.} 
    {\bf a.} Trajectory of an optically trapped particle obtained from a noisy video by DeepTrack (orange) compared to that obtained with the classical radial symmetry algorithm (blue line). Reproduced from~\cite{helgadottir2019digitalDeeptrack}.
    {\bf b.} A U-net can be used to track trapped particles that approach one to the other also when one particle overlaps with the other (defocused particle in the bottom picture on the left). 
    {\bf c.} A TGAN can fill missing frames in a video file (e.g., due to uneven sampling rate) and track the particles allowing the applications of calibration methods that require a constant sampling rate (e.g., those based on power spectral density, autocorrelation functions, and mean squared displacement).
    {\bf d.} An ATN can find the trajectory of optically trapped particles in a video file and use it to determine the physical properties of the particles, such as their refractive index $n_p$ and radius $r$, as well as information about the immersion media, such as its viscosity $\eta$ and its temperature $T$.} 
	\label{fig:tracking}
\end{figure*}

In optical tweezers experiments, particle tracking is a key task.
Deep neural networks have significantly enhanced this task, notably improving the speed and accuracy of detection. Leading tracking algorithms now frequently incorporate Convolutional Neural Networks (CNNs)~\cite{newby2018convolutional,helgadottir2019digital,midtvedt2021quantitative}. These CNNs exhibits greater resistance against noise compared to classical algorithms. This prevents tracking errors due to the presence of noise in the particle video and increases the accuracy of the extracted particle trajectory, as shown in Fig.~\ref{fig:tracking}a. Nevertheless, acquiring enough training data from experiments is challenging because the true values of the position of the trapped particle are not known and may need to be collected manually or with standard methods. To solve this issue, it is possible to train the algorithms on simulated data \cite{helgadottir2019digital,midtvedt2021quantitative}.

An alternative approach that has shown promise is to exploit the symmetries inherent to the tracking problem.
This approach is employed by the recently developed the deep-learning approach called LodeSTAR (Localization and detection from Symmetries, Translations, And 
Rotations)~\cite{midtvedt2022single}.  This approach is particularly beneficial as it enables training on small datasets, even with as little as a single image, without the necessity of ground truth. 
In this way, a single training image is sufficient to train LodeSTAR. 

In addition to the position from images of the particle, deep learning can extract more information, such as the particle's size and orientation. For example, deep learning has been recently used to track the orientation of sperms in an optical trap enabling the extraction of the sperm rotation rates~\cite{OT_rotation_deep_learning2023}. Furthermore, going beyond analyzing images acquired with digital video microscopy, deep learning can potentially be applied also with data acquired with methods based on quadrant-photodiodes (QPDs) or position-sensitive detectors (PSDs). In these cases, deep learning can allow, for example, the extrapolation of the trajectory signal from noisy signals or with frequency higher than the detection bandwidth.

Importantly, deep learning often manage to excel even when standard methods fail. For example, U-nets can be used to track multiple trapped particles that approach one to the other, as schematically illustrated in Fig.~\ref{fig:tracking}b, a situation in which standard methods fail and require complex ad-hoc fixing~\cite{baumgartl2005limits}. This is specially relevant for multiple trapped particles and in case of defocusing (due for example to overlapping of two or many particles).

TGANs could improve the tracking of particles from videos with missing frames or non-constant sampling frequencies thanks to their ability to generate data that respect the temporal correlation of the inputs and thus to generate the missing data from the properties of the phenomenon being studied, as schematically shown in Fig.~\ref{fig:tracking}c. It is possible, for instance, to create a constant sampling rate video from one that is non-constant, enabling the utilization of calibration techniques based on power spectral density, autocorrelation functions, and mean squared displacement.
Instead, ATNs can be used to locate trapped particles in a set of many particles and evaluate their properties (such as dimensions and refractive index) or the fluid properties (such as temperature and viscosity) by identifying how distant points of the trajectory of the particle interact and influence one another, as schematically shown in Fig.~\ref{fig:tracking}d.

\subsection{Trajectory analysis and calibration}

\begin{figure*}[t]
	\centering
	\includegraphics[width=\textwidth]{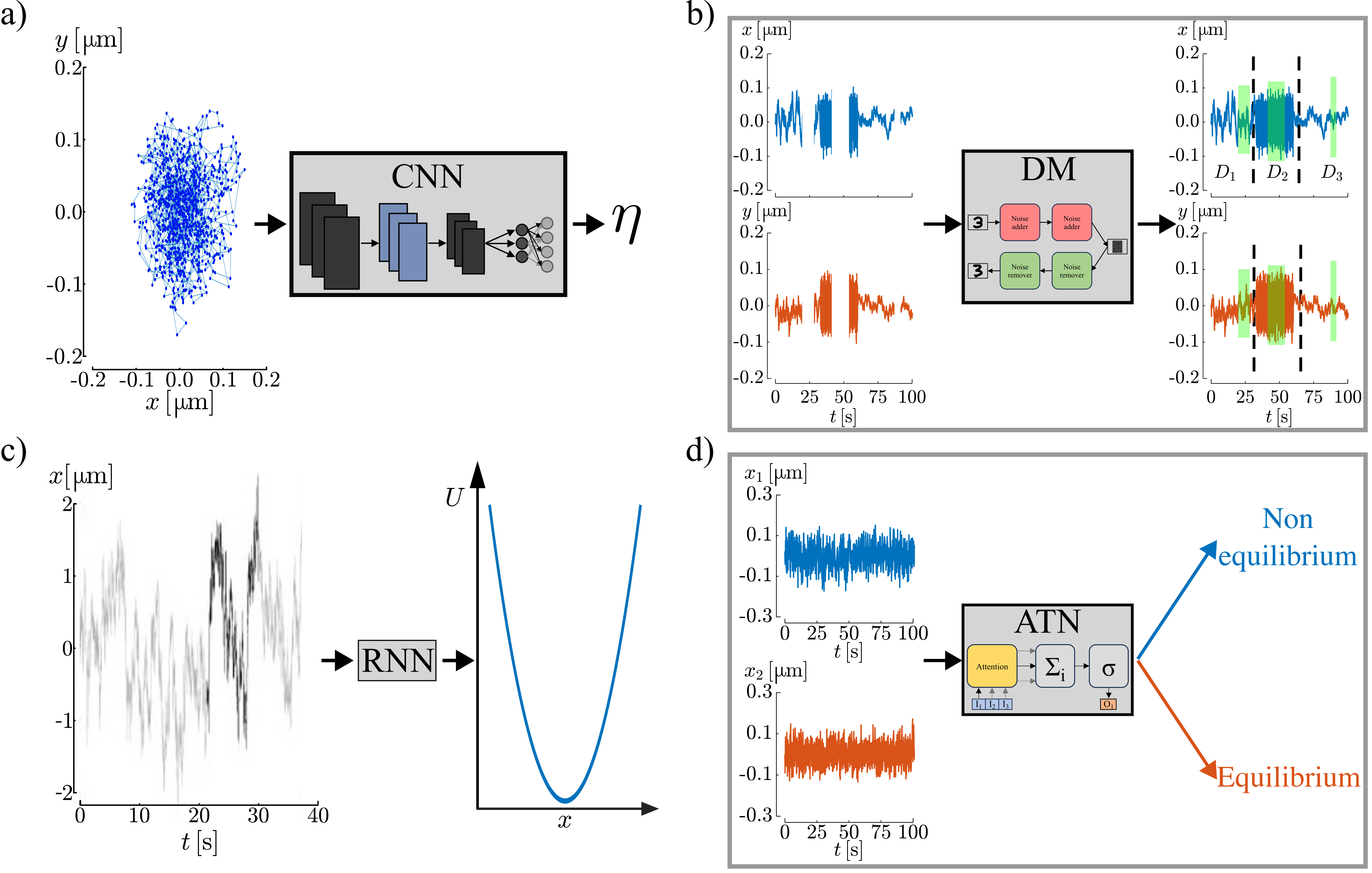}
	\caption{\textbf{Deep Learning for trajectory analysis and calibration.} 
    {\bf a.} A convolutional neural network is trained on simulated data in order to extrapolate from the particle trajectory the medium viscosity $\eta$. Reproduced from~\cite{smith2022machine}.
    {\bf b.} A diffusion model can be used to extract information about the diffusion processes of a trapped particle when there are missing points in the trajectory.
    {\bf c.} The DeepCalib method used a recurrent neural network trained on simulated data to extract the trap stiffness for a microparticle held in a harmonic potential. Reproduced from~\cite{argun2020enhanced}.
    {\bf d.} An attention-based transformer network can determine whether a trapped particle is in thermal equilibrium or in a non equilibrium condition.}
    \label{fig:calibration}
\end{figure*}

Deep learning has proven to be an efficient method for analyzing confined particle motion, especially when experimental conditions change, and has proven effective for calibrating optical tweezers in scenarios where traditional methods are inadequate, such as non-conservative force fields and limited data collection situations.
Recently, the trajectory analysis with deep learning allowed the estimation of rheological properties by reducing the amount of data needed~\cite{smith2022machine}, as schematically shown in Fig.~\ref{fig:calibration}a. This kind of analysis, which would ordinarily require measuring for several minutes, can now be obtained in a matter of seconds. This result was possible by training the neural network on simulated data, further showcasing the potential of synthetic data to be used to train models. In this case, simulating the training data are both essential to get sufficient amounts of data and relatively simple since the equations of motion of a trapped particle are well understood.

Deep learning has also been used to analyze particle trajectories within an optical trap measured using from the forward scattering captured by a quadrant photodiode to discern different kinds of particles~\cite{carvalho2021particle}. 
Potentially, deep learning architectures, such as diffusion models, can be utilized to estimate the properties of various diffusion processes experienced by a trapped particle, even when there are missing points in the trajectory. Indeed, the diffusion model can be employed to reconstruct the particle trajectory by effectively filling in the gaps and can estimate the required properties, as schematically illustrated in Fig.~\ref{fig:calibration}b.

Deep learning can also be used for calibration purposes. This was demonstrated in Ref.~\cite{argun2020enhanced}, where  RNNs were used to estimate force fields with limited data available (trajectory length $<10~\rm{s}$) for harmonic potentials~\cite{argun2020enhanced}, as shown in Fig.~\ref{fig:calibration}c, as well as for more complex and time-varying force fields.
Recent findings underscore the capabilities of neural networks to go beyond determining the stiffness of optical traps, and to estimate properties of trapped particles such as their refractive index or radii~\cite{Hamilton2020}.
The use of deep learning, specifically transformers network, can determine whether a trapped particle is in thermal equilibrium or not, as shown in Fig.~\ref{fig:calibration}d, task that is challenging by using standard methods..

\subsection{Optical force calculations}

\begin{figure*}[t]
    \centering
	\includegraphics[width=\textwidth]{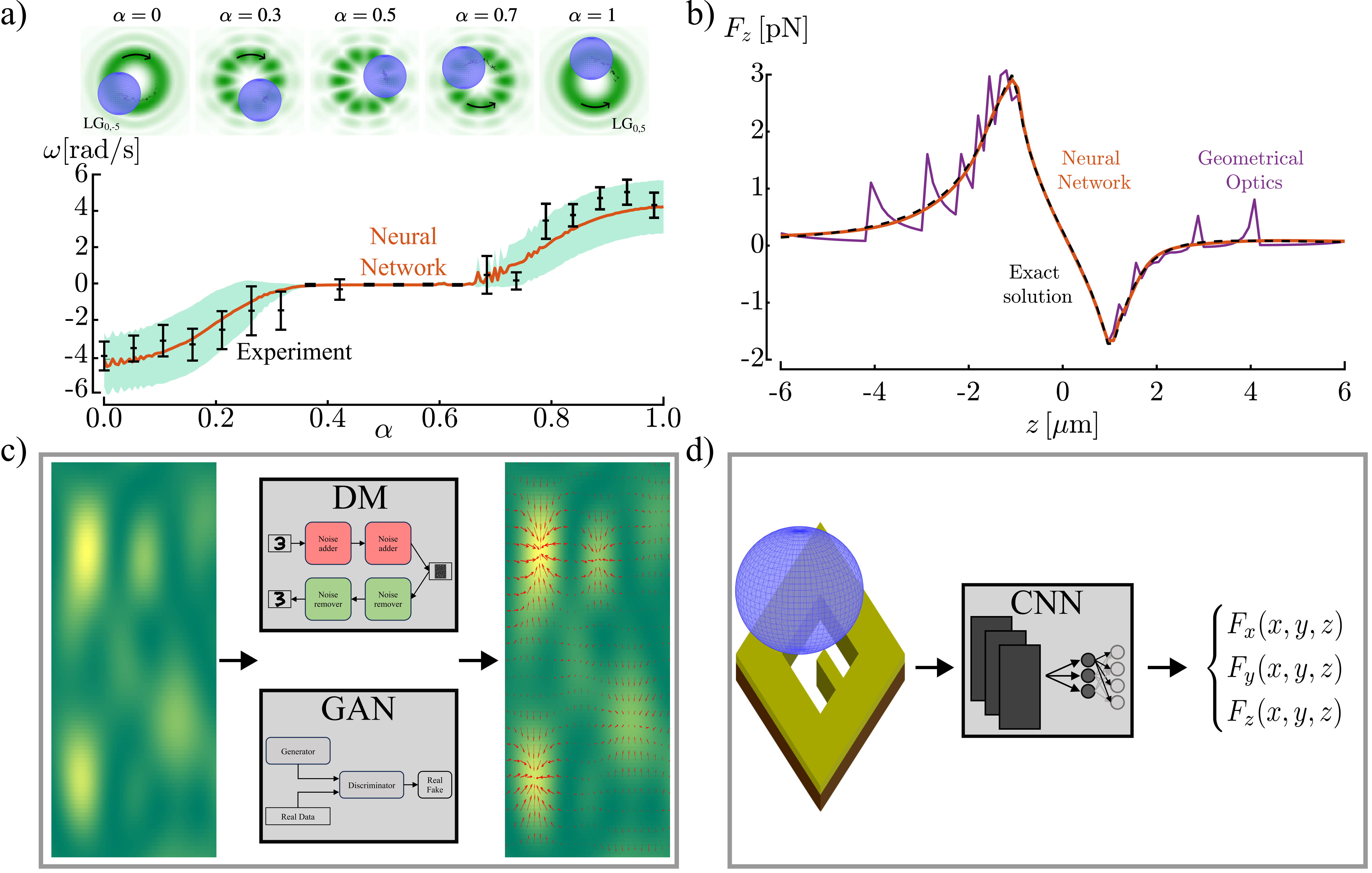}
	\caption{\textbf{Deep learning for optical force calculation.} 
    {\bf a.} Experimental (black symbols) and neural-network-simulated (orange line) rotation rates $\omega$ as a function of the parameter $\alpha$ of the superposition of two Laguerre-Gaussian beams, ${\alpha \, \mathrm{LG}_{0, +5}} + (1-\alpha) \, {\rm LG_{0, -5}}$. The error bars represent standard errors. Reproduced from~\cite{lenton2020machine}.
    {\bf b.} A dense neural network calculates the optical forces in the geometrical-optics approximation increasing not only the calculation speed but also the accuracy when compared to the conventional geometrical-optics approach. The neural network (orange line) has been trained with data generated with geometrical optics using 100 rays (purple line) and approximates much better the exact solution (black line). Reproduced from~\cite{bronte2022faster}.
    {\bf c.} A GNN could evaluate the force field (red arrows in the right panel) directly from images of the optical field (on the left).
    {\bf d.} A CNN could be used to evaluate and optimize the trapping force directly from the 2D design of a near-field optical trap.}
	\label{fig:ForceCalculation}
\end{figure*}

Calculating optical forces can be computationally expensive, especially when optical forces require repeated calculations, such as when simulating the Brownian dynamics of an optically trapped particle~\cite{volpe2013simulation}, or for non-Gaussian beams, such as Laguerre-Gauss beams. Deep learning offers a solution to this problem. For example, neural networks have successfully predicted the forces acting on a spherical trapped particle both in the intermediate regime, even for complex beams~\cite{lenton2020machine}, and in the geometrical-optics approximation~\cite{bronte2022faster}. Importantly, the improvement in speed does not come at the expense of accuracy. Quite the opposite, neural networks have also been shown to be able to overcome some artifacts caused by the restricted number of rays used in the geometrical-optics approximation~\cite{bronte2022faster}. Simple dense neural networks have been shown to perform well for this task, probably thanks to the low dimensionality of both inputs (e.g., the three coordinates od the particle position as well as some of the particle physical properties) and outputs (e.g., the three components of the force). The enhanced computational speed enables simulations of scenarios previously unattainable utilizing conventional computational methods. For instance, modeling a trapped particle that changes size~\cite{lenton2020machine} (Fig.~\ref{fig:ForceCalculation}a), improving the performance and accuracy of geometrical-optics calcualtions~\cite{bronte2022faster} (Fig.~\ref{fig:ForceCalculation}b), exploring the parameter space of an ellipsoid in a double beam configuration~\cite{bronte2022faster}, simulating the dynamics of a trapped red blood cell~\cite{tognato2023modelling}, or evaluating forces produced by beams with amplitude profiles of arbitrary complexity~\cite{malik2022optical}.

As a perspective, DMs and GANs could be used to evaluate the optical forces of complex light fields (also random fields, as speckles field~\cite{evers2013particle,volpe2014brownian,volpe2014speckle,perez2018high}) from intensity images of the field acquired with a camera, as schematically shown in Fig.~\ref{fig:ForceCalculation}c. This is not possible with standard methods, whereas DMs and GANs can learn how an intensity image relates to a force field during the generation process.

Moreover, CNNs, possibly trained with an adversarial approach, could be used to evaluate the optical forces produced by near-field optical trapping from the 2D design of the substrate, as schematically shown in Fig.~\ref{fig:ForceCalculation}d. Currently this design requires the use of numerical methods that requires a lot of computational power and time for having acceptable results.

\subsection{Controlling tweezers}

\begin{figure*}[htp]
	\centering
	\subfloat[]{\includegraphics[width=\textwidth]{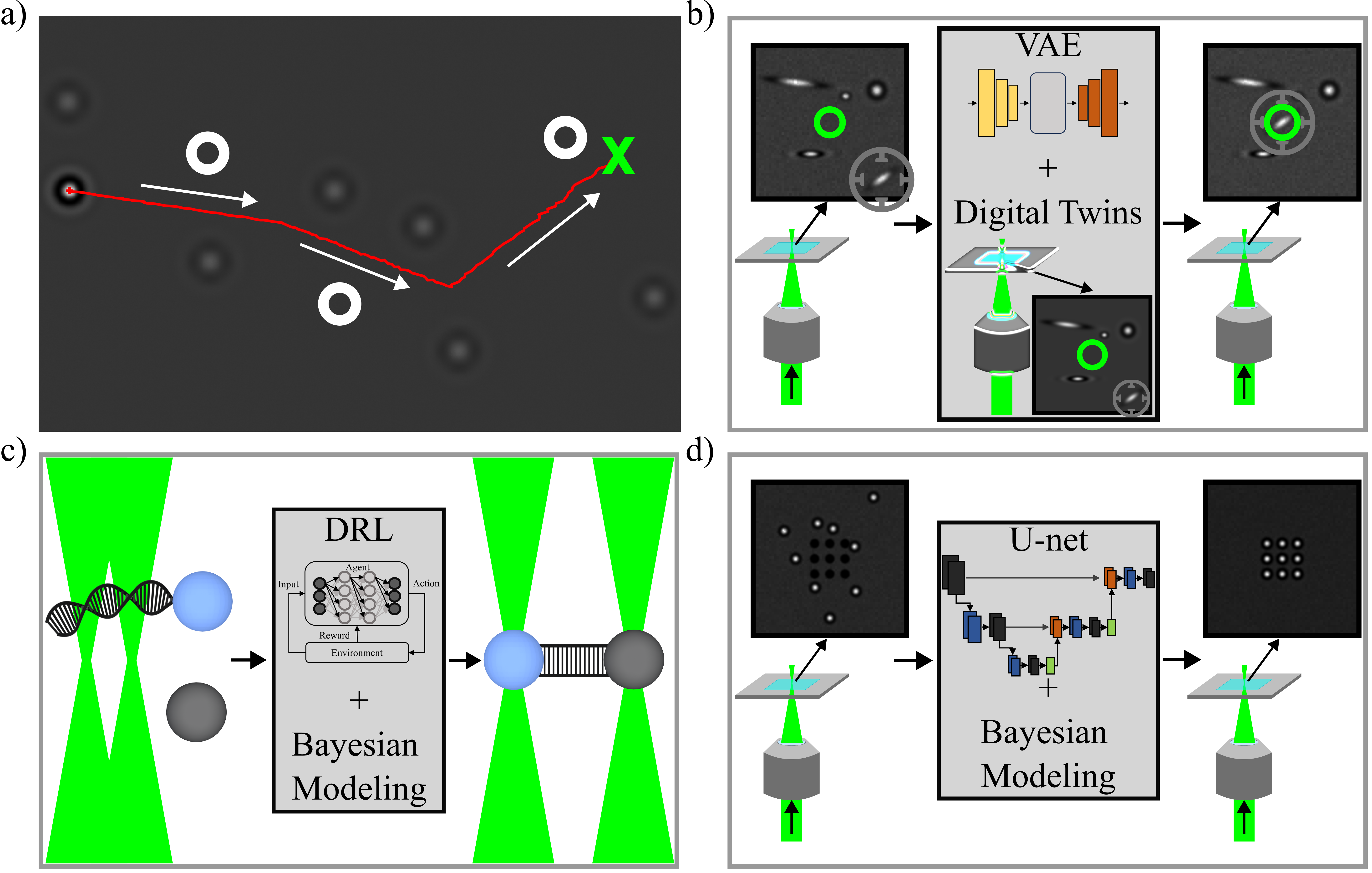}}
	\caption{\textbf{Real-time control of optical tweezers with deep learning.} 
    {\bf a.} Sketch of a trapped particle moved in real time by a neural network to avoid both physical (defocused particles) and virtual (white hollow circles) obstacles. The red solid line represents the trajectory, the white arrows the direction of the motion, and the green cross the destination point of the particle. Reproduced from ~\cite{Praeger2021}. 
    {\bf b.} Digital twins and VAEs can be used to automatize trapping experiments of only particles with specific properties. 
    {\bf c.} Deep reinforcement learning and Bayesian modeling can be used to automatize the DNA pulling experiment done with two optical traps.
    {\bf d.} U-net and Bayesian modeling can improve the process of filling micro-holes in a microfluic chamber with particles in order to create microstructures.}
	\label{fig:controlling}
\end{figure*}

Real-time control of optical tweezers using deep learning can improve their operational efficiency and reliability.
In 2021~\cite{Praeger2021}, a neural network was trained to guide optically trapped particles to precise target positions while avoiding collisions with other particles and obstacles. The first step in this process is to detect particles in images captured by a camera using a thresholding method.
The particle positions are then used to determine the most efficient movements for the captured particle, resulting in its alignment with the desired target. 
This is done by training a deep reinforcement learning algorithm in a simulated environment.
In this way, the NN can determine the most suitable direction for guiding the trapped particle to its target position, all while avoiding potential collisions with other particles, as shown in Fig.~\ref{fig:controlling}a.

To achieve precise optical tweezers control, digital twins can be coupled with deep learning. Digital twins are virtual models of physical objects, systems, or processes, generated by collecting and integrating data from their corresponding physical counterparts~\cite{gelernter1993mirror,grieves2002completing,glaessgen2012digital}. By including optical tweezers within a digital twin framework, researchers can virtually execute and manage microscopic objects, such as individual molecules or nanoparticles, with great precision. This enables improved experimentation at the nanoscale and supplies an abundance of real-time data on the behavior and interactions of the objects. This data can then be analyzed by deep-learning algorithms to optimize experimental conditions and swiftly detect complex patterns and trends that may be difficult for human researchers to discern. For example, digital twins and VAE can be used to automatize trapping experiments of only particles with specific properties as schematically shown in Fig.~\ref{fig:controlling}b. This experiment is not feasible using standard methods because of the need to extrapolate the properties of the particle in real time.

Moreover, Bayesian deep learning can be incorporated into the control structure of optical tweezers to consider possible uncertainties such as sensor noise and variations in particle characteristics. Bayesian deep learning is a deep learning approach using Bayesian modeling, which is a statistical model where the probability is influenced by the belief in the likelihood of a specific outcome~\cite{bayes1763lii}. This, in turn, enables the precise and adaptable manipulation of particles, for example, for drug delivery, for studying biological processes, or for assembling microstructures, as schematically depicted in Figs.~\ref{fig:controlling}c and \ref{fig:controlling}d. The Bayesian framework empowers the system to continuously update its beliefs concerning the state of the particles, thereby enhancing the robustness and efficiency of optical tweezers experiments.

\subsection{Designing optical tweezers} 

\begin{figure*}[htp]
	\centering
	\includegraphics[width=\textwidth]{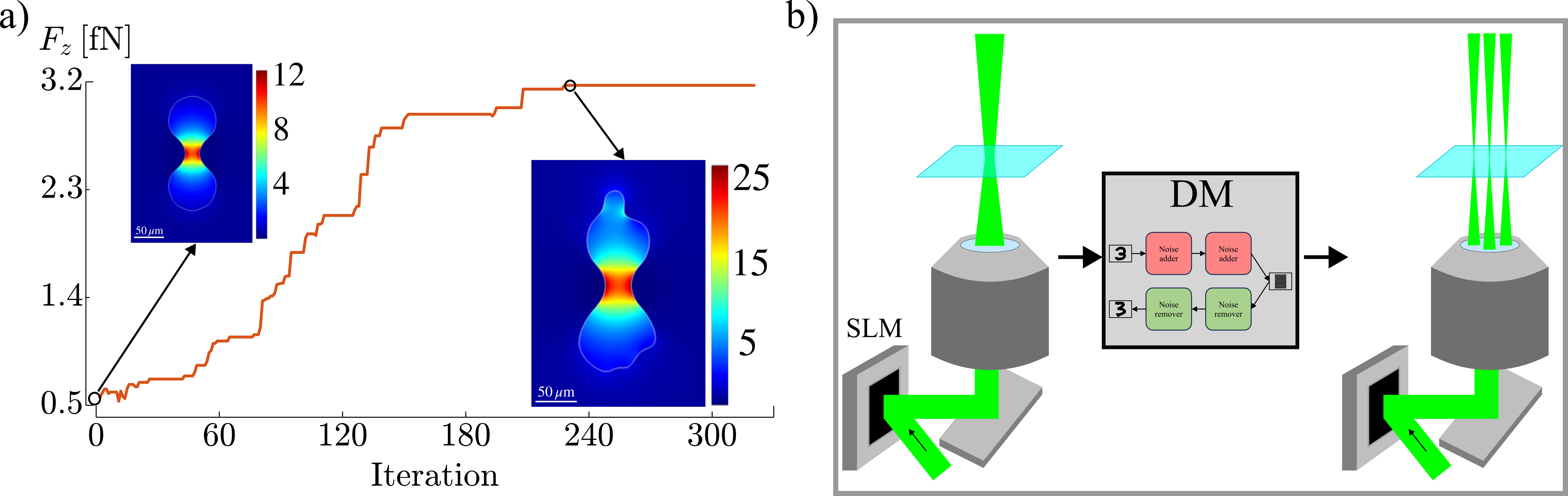}
	\caption{\textbf{Deep learning for designing optical tweezers.} 
    {\bf a.} The design of a nanoaperture is optimized using simulated annealing. The algorithm iteratively updates the shape to find the best one for optical trapping. Reproduced  from~\cite{li2021algorithm}.
    {\bf b.} A diffusion model could be used in combination with a spatial light modulator to trap multiple particles and enhance the focusing, and therefore the trapping force.}
    \label{fig:designing}
\end{figure*}\label{sec:DesigningOT}

Optical tweezers are complex systems whose design can be challenging, especially when using adaptive optics or plasmonic structures. Deep learning has the potential to improve and simplify this design process.
However, until now only probabilistic techniques, such as simulated annealing, have been used to design custom nanostructures that help improve the performance of plasmonic trapping~\cite{li2019algorithmic, li2021algorithm}. By evaluating the optical force produced by different shapes of the nanoaperture, it is possible to optimize its shape, enhance their electromagnetic field, and, therefore, maximize the trapping force, as shown in Fig.~\ref{fig:designing}a.

Deep Learning for designing nanophotonic devices is now widely used~\cite{wiecha2021deep} and its extension for designing optical tweezers is straightforward. 
More advanced techniques, such as deep reinforcement learning combined with digital twins, may improve the design of plasmonic devices. For example, DRL might try different shapes of the nanodevice on the digital twin to find the best shape for the best performance.
Another way to design optical tweezers is to use a spatial light modulator (SLM)~\cite{curtis2002dynamic} and deep learning algorithms to alter the beam shape. Then, the beam shape can be controlled by a diffusion model that generates the appropriate SLM mask, allowing, for example, the trapping of multiple particles with different beam shapes and/or to compensate the spherical aberrations of the optical system, as schematically shown in Fig.~\ref{fig:designing}b.

In addition, digital twins might be used  with VAEs to design the optical elements (e.g., trapping lens properties, laser wavelength) to have specific properties of the optical trap such as a specific stiffness of the trap or a trap able to efficiently trap particles that typically are difficult to trap (e.g., gold nanoparticles, quantum dots, low refractive index particles).

\section{Guidelines}

Considering that many potential applications of deep learning in the optical tweezers domain remain to be developed, we provide here some guidelines.
We also address some specific challenges, such as the availability of only limited datasets and the diversity of optical tweezers setups, which complicate the application of the same techniques broadly to different experiments.

The process of applying deep learning to solve an optical tweezers problem can be broadly split into the following steps: 
1. Problem description. 2. Data collection/simulation. 3. Architecture selection. 4.  Training. 5. Testing.
Often, it is necessary to iterate the process multiple times before achieving an acceptable performance. 

\subsection{Problem description}

The first step in implementing any deep learning model is to provide a detailed description of the problem, outlining what is known and what the deep learning model needs to predict. 
The knowledge of the input and output data, especially which types of data these will contain, is fundamental to choose the proper deep-learning architecture. For instance, the algorithm could use  images  from a camera as inputs and return the commands to send to the laser the beam properties as output.
A key aspect is to define the specific requirements for the sought-after solution. These could be that the output is needed quickly, such as for real-time feedback control, or that the output needs to be accurate, as for image analysis.

When using deep learning to control the experiment, the choice of an architecture able to communicate with the experimental setup and manage the input and output signals is fundamental. A simple solution is to run the deep learning model on a desktop computer connected to the experimental setup. However, more specialized solutions might also be required, for example employing microcontrollers or field programmable gate arrays (FPGAs) with pre-trained neural networks.

Instead, if deep learning is used in data analysis, providing the inputs to the network and retrieving its output is rarely a technical problem. However, it is still recommended to run the algorithm on specialized hardware (GPUs or TPUs), relatively easy and accessible through local computers, servers, or on the cloud.

To enhance the effectiveness and simplify the training of the deep-learning algorithms, the problem needs to be written in as simple terms as possible. For example, the magnitude of the force applied on a sphere in standard optical tweezers depends only on two inputs (radial distance and height from the focus) and not on the three values of the cartesian coordinates ($x,y,z$) because of symmetry arguments.  By exploiting this symmetry in the modelling of the problem, the deep-learning model can perform more accurately and computationally faster, while reducing the requirements of training data and the efforts in training.

Also at the initial stage, it is critical to consider whether deep learning is the best fit for the problem of interest. There are situations in which standard methods perform as well as deep learning with the additional advantage of interpretability and explainability of the results. Instead, a deep-learning model is intrinsically less transparent as it learns through a relatively mysterious training process.
In general, deep learning is preferable  when there is plenty of data for training or when the relation between the inputs and outputs is too complicated to be described analytically or with simple computational models.

\subsection{Data collection/simulation}

Any deep learning approach will require training data to fit the parameters of the model and these data will need to be as representative of the problem as possible.
Depending on the problem at hand and the chosen architecture, the amount of data required for training the neural network will be different. Typically, the quantity of data should be substantial and diverse, representing the entire variable space of the problem. This can easily be the biggest obstacle when applying deep learning. For example, to track the position of a trapped particle, multiple images in different experimental conditions are required to achieve sufficient generality. Nevertheless, some cutting-edge techniques require only a single sample to complete the training, such as the LodeStar tracking algorithm~\cite{midtvedt2022single}.

In several situations, the training data can be produced through simulations allowing access to potentially infinite amounts of data. Multiple software packages help with this, such as DeepTrack~\cite{helgadottir2019digital, midtvedt2021quantitative}, for simulating images of particles, for calculating optical forces~\cite{lenton2020machine,bronte2022faster}, and for analysing trajectories~\cite{argun2020enhanced}. However, the simulated data must be representative of the problem and, to ensure this, a small experimental dataset can be used as a validation set. Sometimes, combinations of simulated and experimental data can improve the learning process. Typically, one would then train the algorithm on the simulated data first and then fine-tune it on the experimental data. 

It is important to highlight that the data should be split into three different subsets: a training set used to train the parameters of the architecture; a validation set used to tune its hyperparameters, i.e. the parameters related to the architecture properties (such as number of neurons, number of layers, dimensions of the layers); and a test set used to evaluate the final performance of the trained model on unseen data (these data should not be used during the optimization of the architecture or the training of the model).

Most algorithms employ supervised learning which requires labelled data. This means that the data must be labeled with the ground truth, i.e., each input of the training dataset needs to be associate to a known desired output that the deep-learning model should provide. Knowing the ground truth is challenging and requires the utilization of standard methods or alternative experimental setups.
There are also unsupervised techniques (e.g., VAEs) that do not need labeled data. In this case, the preparation of the training dataset is much easier a problem, but the validation of the model becomes more challenging and often requires explicit analysis by the user. 

\subsection{Architecture selection}

\begin{table*}[]
    \centering
    \caption{Summary of deep learning algorithms suitable for different problems related to optical tweezers. In the last column, we have listed references that deal with the technique on a general level or apply it in the context of optical trapping or a related field. }
    \begin{tabular}{|l|l|l|}
        \hline
         \textbf{Problem}& \textbf{Model} & \textbf{References}\\ \hline
         
         Particle tracking---single particle 
         & 
         CNNs 
         & 
         \citenum{helgadottir2019digital, midtvedt2021quantitative}\\ 
         \hline
         Particle tracking---multiple particles 
         & 
         U-nets 
         & \citenum{midtvedt2021quantitative} \\ 
         \hline
         Particle classification 
         & 
         CNNs, U-nets 
         &
         \citenum{ronneberger2015u,midtvedt2021quantitative}\\ 
         \hline
         Optical force calculations 
         & 
         DNNs 
         &
         \citenum{lenton2020machine, bronte2022faster} \\ 
         \hline
         Trajectory analysis---single particle 
         & 
         RNNS, ATNs 
         &\citenum{argun2020enhanced} \\ 
         \hline
         Trajectory analysis---multiple particles 
         & 
         GNNs
         & 
         \citenum{pineda2023geometric}\\ 
         \hline
         Calibration 
         & 
         RNNS, ATNs
         &
         \citenum{argun2020enhanced} \\ 
         \hline
         Designing tweezers 
         & 
         Simulated Annealing, VAEs
         &
         \citenum{li2021algorithm,wiecha2021deep} \\ 
         \hline
         Tweezers control---particle movement
         & 
         DRL 
         &
         \citenum{Praeger2021} \\ 
         \hline
    \end{tabular}
    \label{tab:ModelChoice}
\end{table*}

The choice of the architecture to use and its hyperparameters is a crucial point because it greatly influences the performance of the model. To assist with the selection of the appropriate architecture, we have compiled in Table~\ref{tab:ModelChoice} the most commonly utilized architectures for typical tasks relevant to optical trapping and optical manipulation.
The first things to consider are the task to be achieved and the type of data to be analyzed. 

In the case of tracking particles with digital video microscopy, the most commonly used architectures are variants of CNNs. If the goal is to track a single optically trapped particle, a standard CNN is often sufficient \cite{newby2018convolutional,helgadottir2019digital,carvalho2021particle}. However, if many particles need to be tracked simultaneously, then using a U-net is often better than a standard CNN \cite{midtvedt2021quantitative}. 

In the case of trajectory analysis and calibration, an architecture that can handle the time series data is required~\cite{argun2020enhanced,Hamilton2020}. RNNs have been used previously and will often suffice~\cite{argun2020enhanced}. Also, TGANs and ATNs can perform well with various time series and are, therefore, a good option when there are missing data points or complex dependencies in the data. However, if one has a large number of particles that interact, then a GNN is a good choice---as demonstrated by the MAGIK algorithm~\cite{pineda2023geometric}.

To calculate optical forces, DNNs have been shown to work well~\cite{lenton2020machine, bronte2022faster,tognato2023modelling,malik2022optical} and should therefore be the starting point. If the number of input parameters is small (up to a few tens, e.g., the particle position, rotation, and a limited number of parameters describing its shape), then a DNN will almost certainly perform well. Instead, when the number of parameters increases, such as in the case of biological cells which are also deformable, CNNs may be a better choice due to their capacity to capture spatial dependencies and their lower number of fitting parameters.

When deciding on an algorithm to use for controlling optical tweezers, the choice naturally falls on DRL~\cite{Praeger2021}, digital twins, and Bayesian modeling. However, the specific architecture to use is less obvious and depends on the input data. 

Designing optical tweezers with deep learning is an area in which there has not been much research yet, but we believe that generative models, such as GANs and DMs, might be appropriate to deal with the need to generate different designs to find the most efficient one.

There are also cases when one wishes to combine different data types, for example when acquired by different sensors in the same experimental setup. In this case, one option is to use separate models for the different data types, but this restricts the algorithm by not giving the full picture preventing it from investigating correlations between the two different data streams. A superior option is to use hybrid models which combine several architectures. For instance, to handle a time series from a photodiode in combination with images from a camera, one can combine an RNN and a CNN as backbones to make the prediction using a DRL network as a head.

\subsection{Training}

Training consists of adjusting the parameters of a deep learning model to enhance its performance on the specific problem to solve.
It is convenient to use a standard library to implement the models. The two most commonly used are Pytorch~\cite{NEURIPS2019_9015} (which has been on the rise for several years) and Keras/Tensorflow~\cite{tensorflow2015-whitepaper} (which is being slowly abandoned). Often, it is also possible to find already implemented architectures that can be used as a starting point for training your models. For example, the DeepTrack library~\cite{midtvedt2021quantitative,deepTrackRepo} offers an extensive toolkit for image analysis which have been shown to work well on microscopy data. The training process is often computationally demanding, which explains why we recommend running it on specialized hardware (e.g., using a GPU).

Before starting training, it is necessary to select  loss, a performance metric that quantifies how far the model is from the ground truth, providing a quantity to be optimized. 
Therefore, the loss plays a fundamental role during the training process as its value quantifies the ability of the model to predict the real value of the desired parameter accurately.
For example, this can be the square distance between a predicted position and the actual position of a trapped particle, or the proportion of correctly classified samples.

Next, the initialization of the parameters is done, often automatically by the deep-learning framework being employed. Then, the training loop starts. In each iteration, known as an epoch, the training data are split into small batches on which the model is evaluated, and the loss is calculated. The loss is used with an optimization algorithm such as stochastic gradient descent to slightly change the weights of the model to minimize the loss value. Parallel to this, the value of the loss function is calculated on the validation set to see how well the model generalizes its prediction.

Generally, the performance of the model will increase epoch by epoch, but only up to a certain point when measured on the validation set. Afterwards, the validation performance tends to drop due to overfitting. It can be hard to tell for sure if a model is overfitting; generally, the more parameters the model has and the smaller the dataset, the larger the risk of overfitting. To avoid overfitting, it is possible to stop the training when the performance on the validation set has plateaued and before it starts worsening.
Often, tuning of the hyperparameters, such as the number of layers in a CNN, optimizes the results and, also, reduces the risk of overfitting.

\subsection{Testing}

The final step is to test the model to ensure that it performs as desired when applied to new, never-seen-before data. By using as input to the model a validation dataset for which are known the desired outputs, the model output is compared with the expected one. If the performance is satisfactory, then the training process is finished.
If the model has been trained on simulated data, then it is at this stage that the model is tested against real-world data or in an experimental setting. 
However, often the performance is not as good as desired. If the performance on simulated data is significantly better than that on real-world data, this may indicate a discrepancy between the simulations and the experiment.
Similar problems may occur if the training data are experimental but gathered under different  conditions (e.g., a different setup or with a different type of sample). If this happens, it is mandatory to train the model again by using a larger or more representative training dataset.

When employing the model in a real-time experimental setting, there is often a need for the model to make its predictions quickly. To achieve the required computational speed (especially when using the model in embedded systems, such as microcontrollers or FPGAs), connections or entire neurons may be removed from the neural network to reduce the size and increase the speed. This operation is called pruning. The aim is to strike an optimal trade-off between speed and accuracy for a real-time application and this requires further testing.

\section{Conclusions}

In this perspective, we investigate the application of deep learning for the optical tweezers field.
As examples, we discuss the improvements in particle tracking at low signal-to-noise levels~\cite{helgadottir2019digitalDeeptrack} and in quantifying the rotation of trapped particles~\cite{OT_rotation_deep_learning2023}. 
Furthermore, we highlight the use of deep learning to address cases that traditional methods cannot deal with, such as accurately tracking multiple particles when they are close together, filling in missed frames in videos, or selectively tracking particles with unique characteristics.

Then, we discuss the enhancement of trajectory analysis and optical tweezers calibration, which permit one to estimate rheological properties with only a few seconds of data instead of minutes~\cite{smith2022machine} and, also, to discern different typologies of particles~\cite{carvalho2021particle}. Moreover, we propose to use deep learning in some cases when standard methods fail:
DMs may help to reconstruct trajectories with missing data points and estimate the desired properties; ATNs may help to determine whether a trapped particle is in thermal equilibrium or not.

Furthermore, deep learning has already improved the calculation of optical forces by increasing the computational speed and accuracy~\cite{bronte2022faster} and by allowing the study of nontrivial cases (such as with Laguerre-Gaussian beams)~\cite{lenton2020machine}. In this scenario, optical forces could be calculated also in cases where standard methods are not viable. Indeed, DMs and GANs can calculate the force field starting from intensity images of the optical field, while CNNs can do the same from the design of a near-field optical trap.

When real-time control of optical tweezers is necessary, standard methods are often too computationally slow. Recently, NNs have allowed moving a trapped particle to a target position while avoiding collisions with real and virtual obstacles~\cite{Praeger2021}. We believe that real-time control and automatization of optical tweezers can be further improved using deep learning. Digital twins with VAEs may be suitable when the automatic trapping of specific particles is desired. DRL with Bayesian modeling may automate experiments, such as DNA pulling, optimizing the search for favorable experimental conditions. U-nets with Bayesian modeling may help automate the process of designing microstructures with optimal optical manipulation properties. 

Designing optical tweezers can be challenging, especially when more complex designs are required. Deep learning can provide an effective solution for these requirements. Although the design of optical tweezers has thus far only utilized probabilistic methods like simulated annealing~\cite{li2021algorithm}, deep learning has the potential to enhance this process. For example, DMs can design optical tweezers with a spatial light modulator for trapping multiple particles while enhancing the trapping force.

Finally, we provide guidelines for using deep learning in optical trapping and optical manipulation, highlighting step-by-step the process to follow to create an effective deep learning model, from the problem description to the model validation, while avoiding common pitfalls.

\begin{acknowledgments}
We acknowledge support from the MSCA-ITN-ETN project ActiveMatter sponsored by the European Commission (Horizon 2020, Project No. 812780), the Horizon Europe ERC Consolidator Grant MAPEI (grant number 101001267), and the Knut and Alice Wallenberg Foundation (grant number 2019.0079).
We would also like to thank Agnese Callegari and Caroline B. Adiels for their helpful discussions about the format of this article, which significantly improved the final manuscript.
\end{acknowledgments}

\end{document}